\documentclass{agn_2000}
\usepackage{graphics}
\begin{document}
\title{The Central Regions of Early-Type Galaxies Hosting Active Galactic
Nuclei as viewed with {\it HST}/NICMOS}
\author{Swara Ravindranath\inst{1,2}, Luis C. Ho\inst{1}, C. Y. Peng\inst{3},
A. V. Filippenko\inst{2}, \and W. L. W. Sargent\inst{4}}  
\institute{The Observatories of the Carnegie Institution of Washington, 813 Santa 
Barbara Street, Pasadena, CA 91101
\and Astronomy Department, University of California, Berkeley, CA 94720$-$3411
\and Steward Observatory, University of Arizona, Tucson, AZ 85721
\and Palomar Observatory, Caltech 105$-$24, Pasadena, CA 91125}
\titlerunning{AGN Host Galaxies}
\maketitle

\begin{abstract}

We present preliminary results from surface photometry of a sample of
early--type galaxies observed with NICMOS on board {\it HST} in
the F160W ($H$-band) filter.
Dust is found to occur mostly in galaxies which possess unresolved nuclei,
and in a few cases the presence of dust causes the isophotes to appear boxy.
We performed a 2--D modeling of the host galaxy light 
distribution using the empirical Nuker function, which proves to be an 
adequate representation of the surface brightness profile for most early--type 
galaxies. The accurate modeling of the bulge light allows us to obtain the
magnitude of the unresolved central source without contamination from the
underlying stellar population. Our classification of early--type galaxies into
core--type and power--law galaxies agrees with the results from previous work
based on {\it HST} optical images, with core profiles occurring in luminous 
galaxies and power--law profiles occurring in low--luminosity galaxies with 
disky isophotes. We do not find a dichotomy in the distribution of inner
slopes. Unresolved nuclear point sources are evident in about 
one--third of the sample galaxies, with magnitudes in the range
$13.5 \leq m_{H} \leq 17.3$. Most of the type 1 Seyfert and LINER nuclei
(83\%) reveal a distinct point source, while the detection rate is much less
for type 2 nuclei (35\%). The occurrence of point sources appears to be 
independent of whether the underlying galaxy has core--type profile or 
power--law profile.  

\end{abstract} 

\section{Introduction}

Images of the central regions of galaxies at {\it HST} resolution have made
it possible to probe structures very close to the nucleus and to reveal 
unresolved point sources or compact star clusters. Optical images obtained
using {\it HST} showed that the King model and the de Vaucouleurs $r^{1/4}$
law which were conventionally used to describe the surface brightness profiles
of early--type galaxies do not provide a good fit for the regions within
$\sim$ 10$^{\prime\prime}$ (Ferrarese et al. 1994; Lauer et al. 1995;
Carollo et al. 1997; Faber et al. 1997). 
The surface brightness profiles in the central regions can be
parameterized by the empirical ``Nuker'' function, which is essentially a
double power--law of the form
\begin{equation}
I(r)=2^{(\beta-\gamma)/\alpha}I_{b}\left(\frac{r_{b}}{r}\right)^{\gamma}
\left[1+\left(\frac{r}{r_{b}}\right)^{\alpha}\right]^{(\gamma-\beta)/\alpha},
\end{equation}
where $\beta$ is the slope outside the break radius $r_{b}$, $\gamma$ is the
inner slope, $\alpha$ controls the sharpness of the transition between the 
outer and inner slopes, and $I_{b}$ is the surface brightness at the break 
radius (Lauer et al. 1995). Early--type galaxies can be classified
into core--type or power--law systems based on their surface brightness
profiles. Core--type galaxies show a significant change in the slope of the 
surface brightness profile within the break radius $r_{b}$ with the inner slope
$\gamma$ being much shallower than for the outer regions. Power--law galaxies
have profiles that show no significant change in slope within the inner 
10$^{\prime\prime}$ and remain as steep power laws all the way to the resolution
limit ($r=0\farcs1$). A value of $\gamma$ $<$ 0.3 is adopted as the 
criterion for classifying a galaxy as a core--type. Faber et al. (1997) 
found that 
no galaxies had $\gamma$ in the range 0.3 to 0.5 and therefore that there was a 
clear dichotomy in the distribution of $\gamma$ values between core galaxies
and power--law galaxies. Interestingly, the surface brightness profiles
derived from WFPC2 $V$--band images of early--type galaxies with kinematically
distinct cores yielded values of $\gamma$ between 0.3 and 0.5, thereby 
removing the dichotomy (Carollo et al. 1997). Surface photometry of early--type
galaxies based on $H$--band images from {\it HST} have been presented by
Quillen, Bower, \& Stritzinger (2000), along with the analytic Nuker fits,
but they did not decouple the contributions from the bulge and point sources.
 
In this work, we aim to model the galaxy surface brightness closely 
and to extract the point source magnitudes free from contamination
by the bulge light, for a sample of galaxies that host AGNs. Parameterizing
the bulge and nucleus as separate components enables us to probe the relationship
between the properties of the host galaxy and the nuclear activity. The NIR
images are most suitable for deriving the surface brightness profiles
because of 
the reduced contamination from dust and young star clusters, and they sample 
the underlying smooth stellar population better than optical images.

\begin{table}
      \caption{Properties of sample galaxies$^{\rm a}$}
         \label{KapSou}
      \[
         \begin{array}{p{0.2\linewidth}lccc}
            \hline
            \noalign{\smallskip}
  Galaxy&{\rm Hubble\ Type}&{\rm {\it B_{T}}}&{\rm {\it M_{B}^{o}}}&{\rm Spectral\ Class} \\
            \noalign{\smallskip}
            \hline
            \noalign{\smallskip}
            NGC 221&{\rm E2}&9.0&-15.5&{\rm A} \\
            NGC 404&{\rm S0}&11.2&-15.9&{\rm L2}\\
            NGC 474&{\rm S0}&12.3&-20.5&{\rm L2::}\\
            NGC 524&{\rm S0}&11.3&-21.3&{\rm T2::}\\
            NGC 821&{\rm E6?}&11.6&-20.1&{\rm A}\\
            NGC 1052&{\rm E4}&11.4&-19.9&{\rm L1.9}\\
            NGC 2685&{\rm SB0}&12.1&-19.2&{\rm S2/T2}\\
            NGC 3115&{\rm S0}&9.8&-19.3&{\rm A}\\
            NGC 3379&{\rm E1}&10.2&-19.3&{\rm L2/T2::}\\
            NGC 3384&{\rm SB0}&10.8&-18.7&{\rm A}\\
            NGC 3593&{\rm S0/a}&11.8&-17.2&{\rm H}\\
            NGC 3900&{\rm S0}&12.2&-20.1&{\rm L2:}\\
            NGC 4026&{\rm S0}&11.6&-19.5&{\rm A}\\
            NGC 4111&{\rm S0~spin}&11.6&-19.5&{\rm L2}\\
            NGC 4143&{\rm SAB0}&11.6&-19.2&{\rm L1.9}\\
            NGC 4150&{\rm S0}&12.4&-17.5&{\rm T2}\\
            NGC 4261&{\rm E2}&11.4&-21.3&{\rm L2}\\
            NGC 4278&{\rm E1}&11.0&-18.9&{\rm L1.9}\\
            NGC 4291&{\rm E}&12.4&-20.0&{\rm A}\\
            NGC 4374&{\rm E1}&10.0&-21.1&{\rm L2}\\
            NGC 4406&{\rm E3}&9.8&-21.3&{\rm A}\\
            NGC 4417&{\rm SB0:spin}&12.0&-19.1&{\rm A}\\
            NGC 4472&{\rm E2}&9.3&-21.8&{\rm S2::}\\
            NGC 4589&{\rm E2}&11.6&-20.7&{\rm L2}\\
            NGC 4636&{\rm E0}&10.4&-20.7&{\rm L1.9}\\
            NGC 5273&{\rm S0}&12.4&-19.2&{\rm S1.5}\\
            NGC 5548&{\rm S0/a}&13.3&-21.3&{\rm S1.5}\\
            NGC 5838&{\rm S0}&11.9&-20.5&{\rm T2::}\\
            NGC 5982&{\rm E3}&12.0&-20.8&{\rm L2::}\\
            NGC 6340&{\rm S0/a}&11.8&-20.0&{\rm L2}\\
            NGC 7457&{\rm S0}&12.0&-18.6&{\rm A}\\
            NGC 7626&{\rm E:pec}&12.1&-21.2&{\rm L2::}\\
            NGC 7743&{\rm SB0}&12.3&-19.7&{\rm S2}\\ 
            \noalign{\smallskip}
             \hline
         \end{array}
      \]
\begin{list}{}{}
\item[$^{\rm a}$] Galaxy properties taken from Ho et al. (1997) 
\end{list}
   \end{table} 

\begin{figure}
 \resizebox{\hsize}{!}{\includegraphics{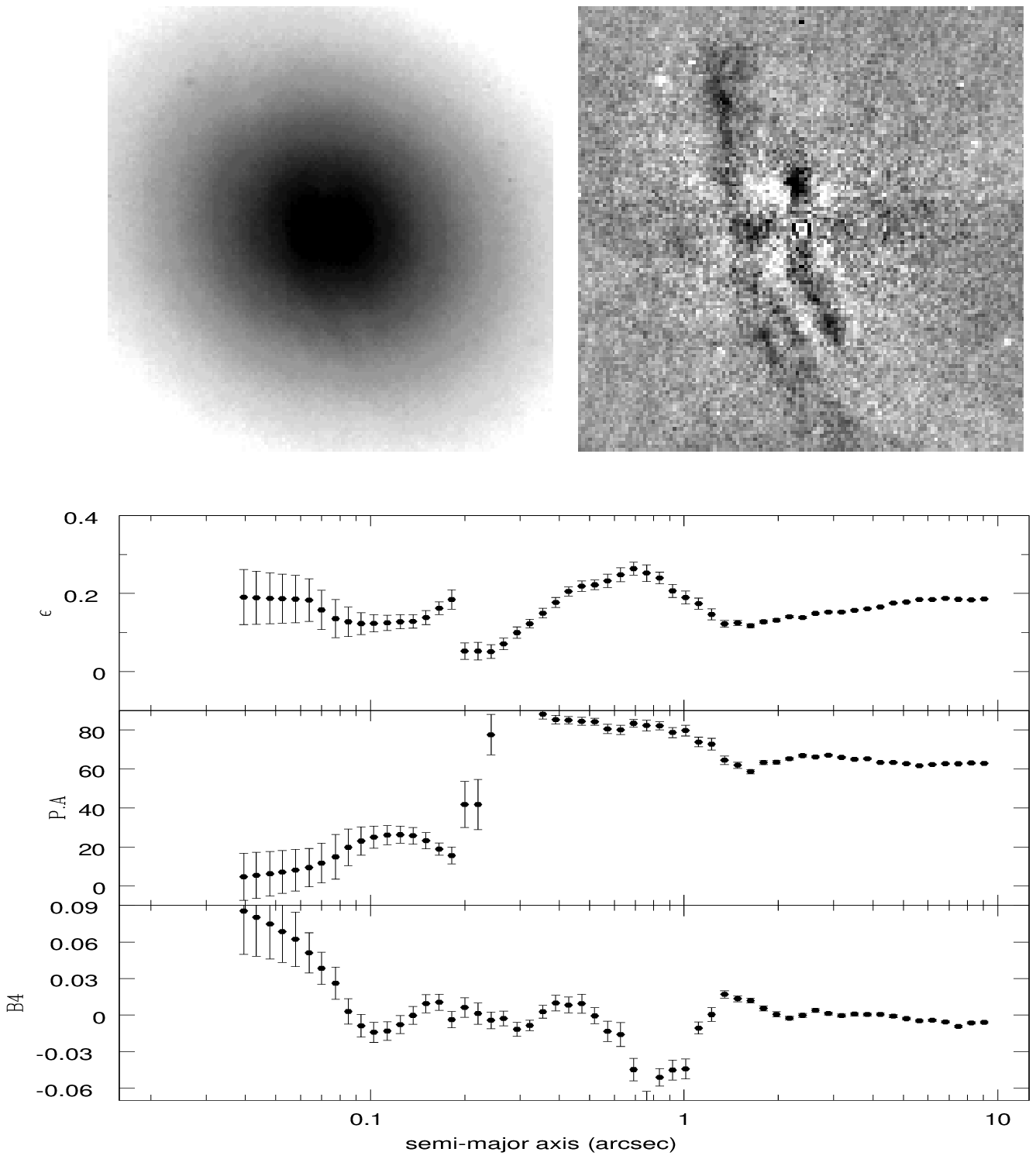}}
 \vskip 0.0cm
 \resizebox{\hsize}{!}{\includegraphics{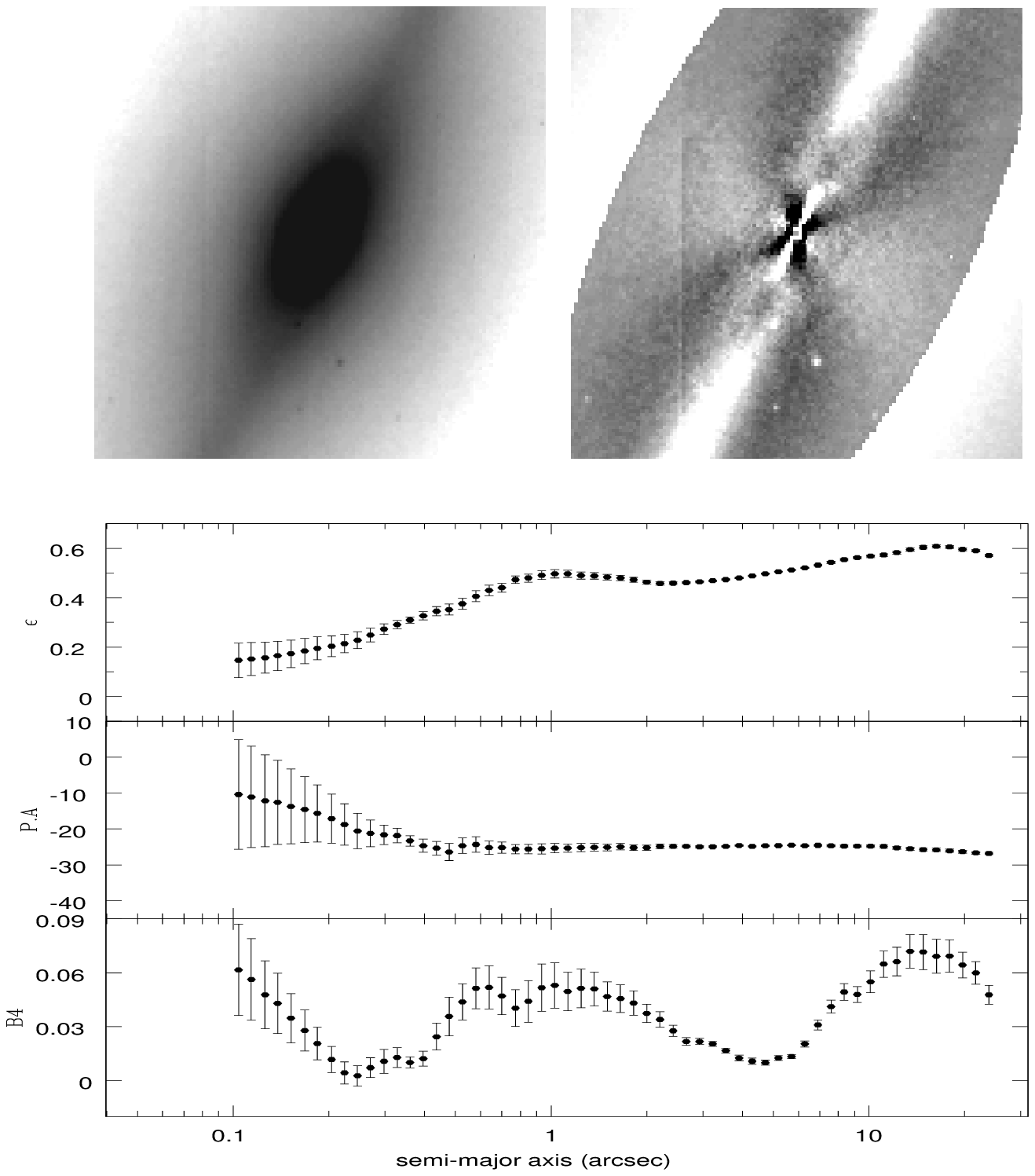}}
\caption[]{The results of isophotal analysis are shown for NGC 4374 ({\it top})
and NGC 3115 ({\it bottom}). The left image is the observed $H$--band image
and the right image is the residual after subtracting the model derived from
elliptical isophotes. Each image is 10$^{\prime\prime}$ $\times$ 10$^{\prime\prime}$
and is centered on the galaxy. The orientation is arbitrary. The lower panels show the
variation of ellipticity, position angle, and the shape parameter (B4) along the
semi--major axis.}
\end{figure}                                                                          
\section{The Sample and Data Analysis}

Early-type (E and S0) galaxies were selected from the sample of galaxies (Table 1)
studied spectroscopically by Ho, Filippenko, \& Sargent (1997). 
Archival {\it HST} NICMOS
images are available for 33 of these galaxies (14 E, 16 S0, and
3 S0/a) from different snapshot survey programs. The data were
obtained in the F160W filter using NIC2 and NIC3 cameras. 
The NIC2 images
have a field of view of 19$\farcs$2 $\times$ 19$\farcs$2 with
a scale of 0$\farcs$076/pixel. The NIC3 images have a field of view of 
51$\farcs$2 $\times$ 51$\farcs$2 with a scale of 
0$\farcs$2/pixel.
The CalnicA images were used after performing additional steps
suggested in the NICMOS handbook which include pedestal removal, creating 
bad pixel masks, and 
interpolating over the bad pixels (Dickinson 1999). The ramp fitting procedure used in 
the CalnicA removes most of the cosmic rays from the images.
Residual cosmic rays, defective pixels, and the coronographic hole (in
NIC2 images) were masked during further analysis.

\begin{figure}
 \resizebox{\hsize}{7cm}{\includegraphics{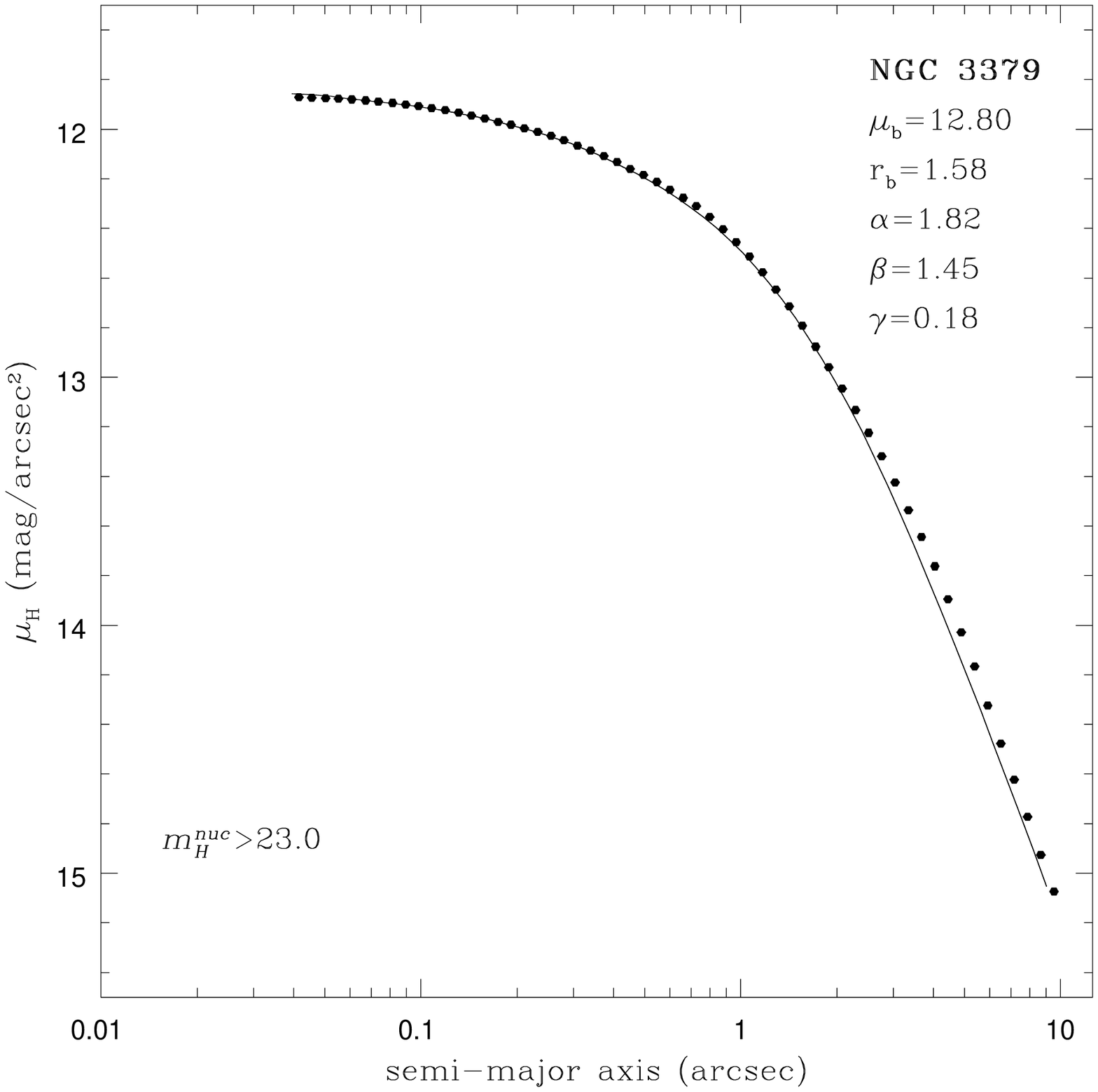}}
  \resizebox{\hsize}{7cm}{\includegraphics{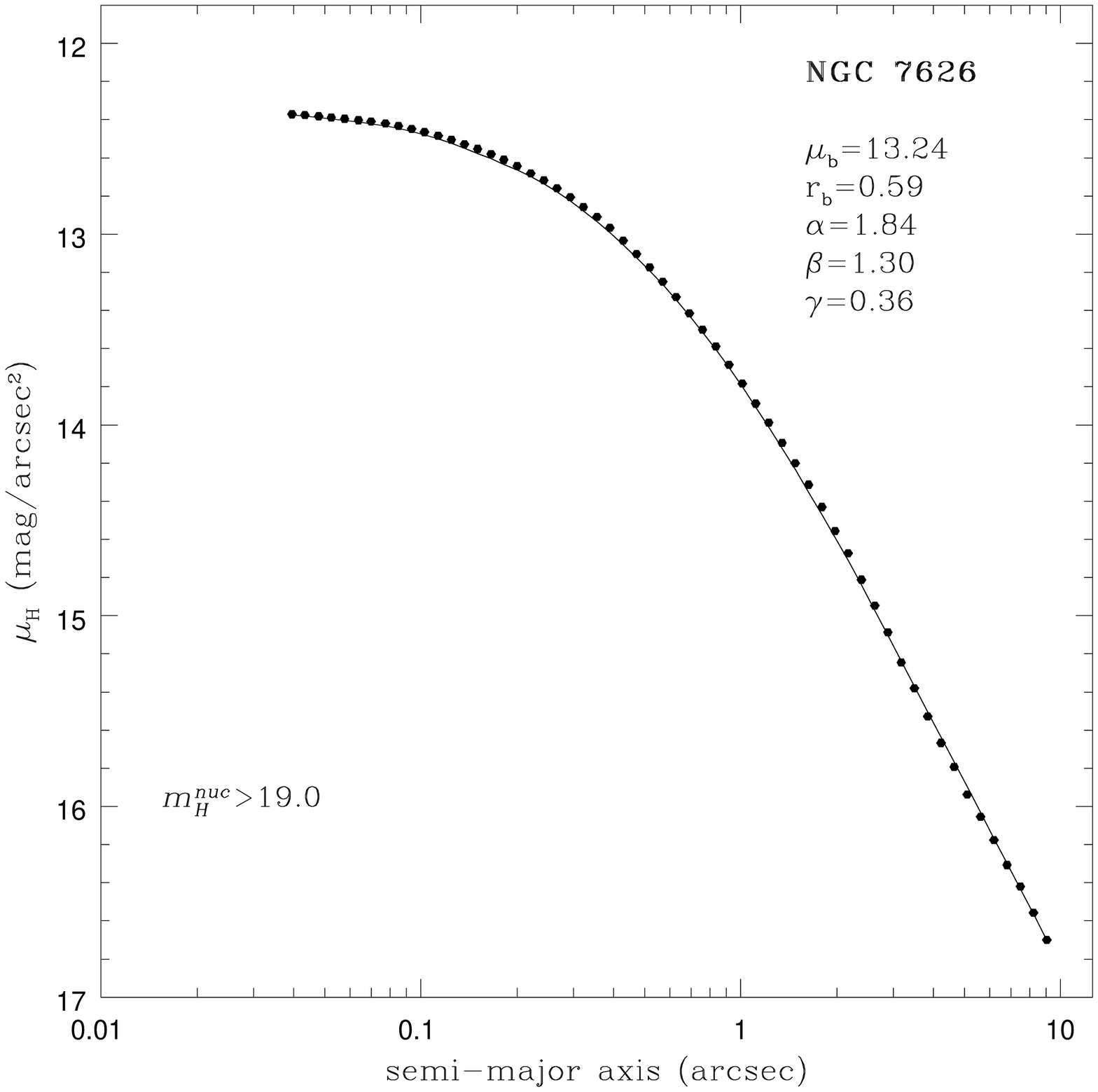}}
\caption[]{Nuker--law fits to the core--type galaxy NGC 3379 ({\it top})
and the power--law galaxy NGC 7626 ({\it bottom}). The solid points
represent the observed surface brightness profile. The solid line is the
1--D profile extracted from isophote fits to the generated model galaxy.
Upper limits to nuclear magnitudes are given.}
\end{figure}

\section{Isophotal Analysis}

The {\it ellipse} task available in STSDAS was used to fit
elliptical isophotes to the galaxy images and thereby obtain the surface
brightness profiles and the radial variation of ellipticity and position
angle. The results of the isophotal analysis were used to create smooth
model images of the galaxies, and residual images which highlight the 
high--spatial frequency features were obtained by subtracting the model images
from the observed images. The presence of features such as dust and weak disks 
causes variations in ellipticity and P.A. The deviation from perfect elliptical
isophotes can be expressed as a Fourier series, and the Fourier coefficient 
of the fourth cosine term in the expansion series (B4) 
is negative for boxy isophotes and positive for disky isophotes
(Jedrzejewski 1987). 
The results from the isophotal analysis are shown in Figure 1 for NGC 4374, in which
dust is revealed in the residual image, and for NGC 3115, which has a dominant stellar
disk. Notice that the isophotes tend to get boxy around $r=0$\farcs$2
-0$\farcs$8$ in the case of NGC 4374 due to the dust lanes.

\begin{figure}
 \resizebox{\hsize}{7cm}{\includegraphics{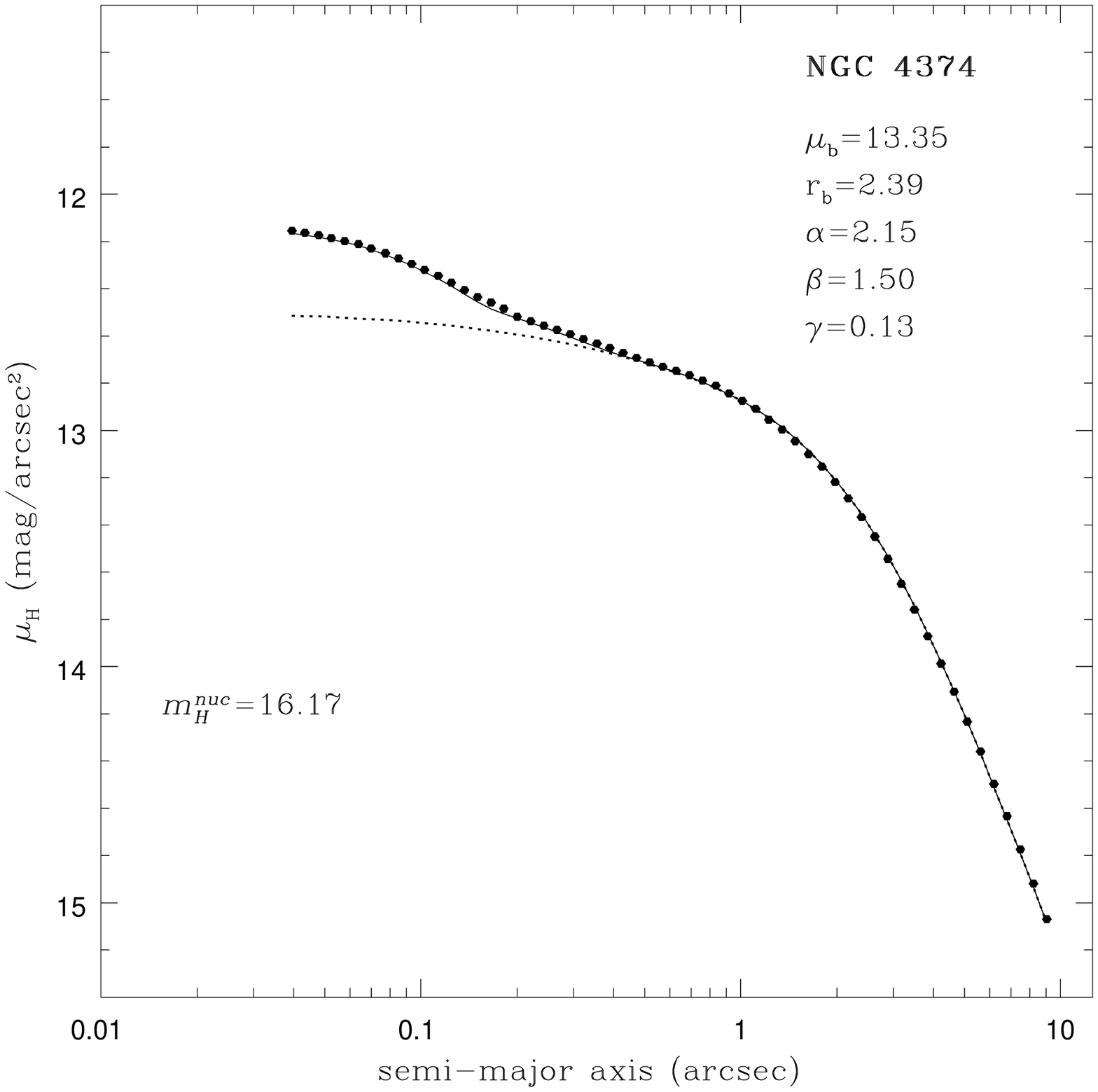}}
  \resizebox{\hsize}{7cm}{\includegraphics{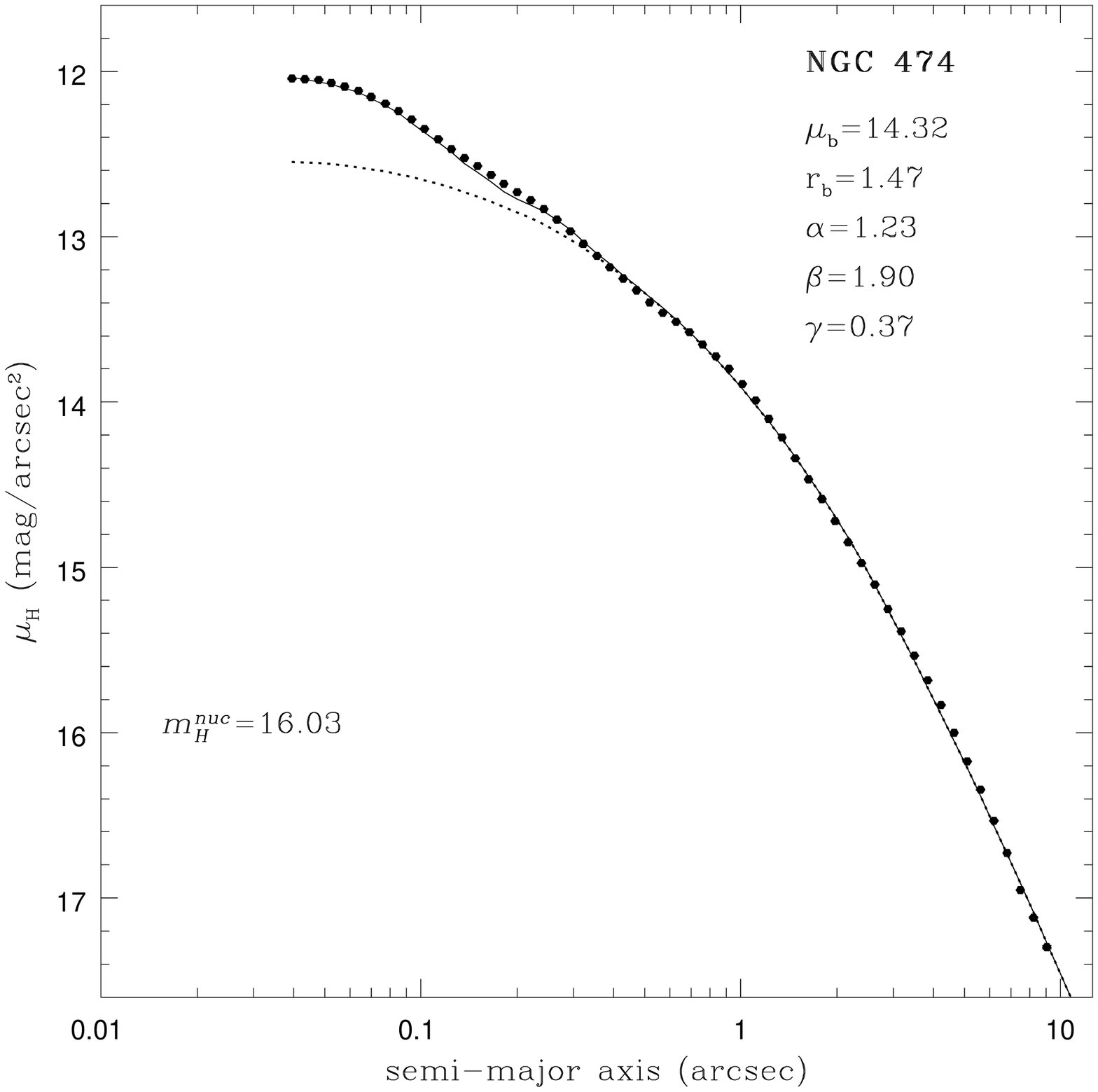}}
\caption[]{Nuker--law fits with point source for the core--type galaxy NGC 4374 
({\it top})
and the power--law galaxy NGC 474 ({\it bottom}). The solid points
represent the observed surface brightness profile. The solid line is the
1--D profile extracted from isophote fits to the generated model galaxy.
The dotted line is the intrinsic bulge profile obtained
by omitting the point source in the model.}
\end{figure}

\section{2--D Modeling Using Analytic Functions}

Earlier {\it HST} studies of the central surface brightness profiles of 
galaxies involved deriving the central parameters by fitting the 
1--D profiles with analytic functions. The uncertainities caused
by the PSF were overcome by performing the fits on 
profiles derived from deconvolved images (Lauer et al. 1995) or by
convolving the analytic 1--D fits with the PSF and then comparing to the 
observed profiles (Carollo \& Stiavelli 1998). 
Most of the data used in the present work were obtained through
snapshot survey programs and do not have the high S/N required to perform
reliable deconvolution of the images. Since a more reliable approach would be
to sample the PSF in two dimensions, we choose to parameterize the galaxy bulge 
through a 2--D modeling using analytic 
functions and convolving the model image with the 2-D PSF image.
This was done using the GALFIT program (Peng et al., in preparation), which 
enables us to create model
images for different analytic functions (e.g., S\'{e}rsic/de Vaucouleurs, 
Nuker, exponential, Gaussian) and convolve them with the PSF. 
In lieu of observing PSFs simultaneously, TINYTIM simulations provide a
good alternative (Krist \& Hook 1997). 
GALFIT allows modeling the contribution from unresolved
point sources or resolved star clusters in addition to the underlying galaxy.
The program outputs the fit parameters for the galaxy and the point source
or star cluster, along with images of the model galaxy and residuals.
Following the results from earlier {\it HST} studies of early-type
galaxies, we modeled the galaxy light using the empirical Nuker function.
When appropriate, we also fit a compact source simultaneously in order
to deblend a component in excess of the underlying galaxy component.
When no compact source was required, we obtained an upper limit to the
point source magnitude.
A comparison of the 1-D surface brightness profiles derived by doing
ellipse--fits on the observed image and the model image created by GALFIT, 
shows that the 2-D modeling reproduces the observed profiles very 
well (Figures 2 and 3).  
 
\section{Results}

\subsection{Properties of the Host Galaxies}

Early--type galaxies appear to have a smooth light distribution in the 
$H$ band,
showing dust features only in very few cases, mostly associated with galaxies
hosting unresolved nuclei (e.g., NGC 1052, 4150, and 4374). The presence of
dust in some cases is found to cause boxiness of the isophotes resulting in 
negative B4 values. We confirm the results from earlier studies that core galaxies
are luminous elliptical galaxies with boxy or neutral isophotes, while 
power--law galaxies have low luminosities and disky isophotes (Faber et al. 1997). 
However, we do not find a dichotomy in the inner slope values
(Ravindranath et al. 2001). Instead, we find a 
continuous distribution of $\gamma$ values similar to that found by Carollo et al (1997)
for their sample of ellipticals with kinematically distinct cores. Core galaxies
obey the fundamental--plane (FP) relations in the (log $r_{b}$, $\mu_{b}$)--plane with 
brighter galaxies having larger cores with lower surface brightness (Figure 4). 

\begin{figure}
  \resizebox{\hsize}{!}{\includegraphics{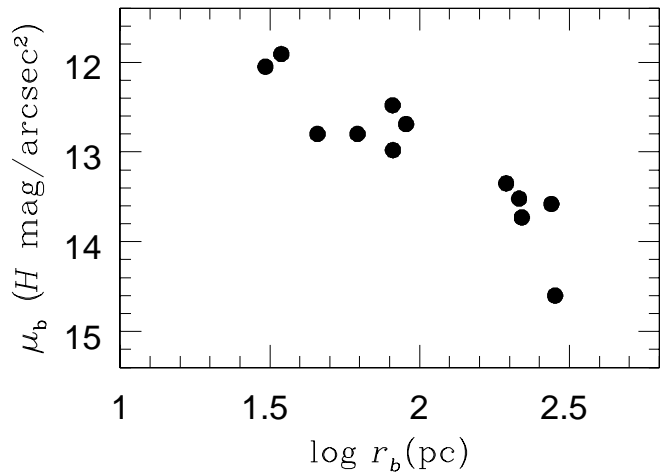}}
\caption[]{The FP relation between $\mu_{b}$ and $r_{b}$
for core galaxies.}
\end{figure}

\subsection{Properties of Unresolved Nuclei}

About one-third of our sample galaxies have clear evidence for the presence of 
an unresolved
central compact source, and the frequency of occurrence of point sources is almost 
equal for core galaxies and power-law galaxies. However, all the central sources 
associated with core galaxies are AGNs by their nuclear spectral classification, 
while power-law galaxies mostly contain absorption--line nuclei, 
transition objects, 
and weak AGNs. Most type 1 Seyfert and LINER nuclei (5 out of 6) have point sources 
compared to type 2 nuclei (5 out of 14). 
The unresolved nuclei have magnitudes in the range $13.5 \leq m_{H} \leq 17.3$,
with the exceptionally bright Seyfert nucleus in NGC 5548 having $m_{H}=11.9$ mag.

\end{document}